\documentclass[aps,prx,twocolumn,showpacs,superscriptaddress]{revtex4-1}

\usepackage{graphics}
\usepackage{graphicx}
\usepackage{dcolumn}
\usepackage{bm}
\usepackage{color}
\usepackage{amssymb,amsmath}

\begin{document}

\title{Anderson localisation and optical-event horizons in  rogue-soliton generation}
\author{Mohammed F. Saleh}
\affiliation{Scottish Universities Physics Alliance (SUPA), Institute of Photonics and Quantum Sciences, Heriot-Watt University, EH14 4AS Edinburgh, UK}
\affiliation{Department of Mathematics and Engineering Physics, Alexandria University, Alexandria, Egypt}

\author{Claudio Conti}
\affiliation{Department of Physics, University Sapienza, Piazzale Aldo Moro 2, 00185 Rome, Italy}
\affiliation{Institute for Complex Systems (ISC-CNR),  Via dei Taurini 19, 00185 Rome, Italy}

\author{Fabio Biancalana}
\affiliation{Scottish Universities Physics Alliance (SUPA), Institute of Photonics and Quantum Sciences, Heriot-Watt University, EH14 4AS Edinburgh, UK}
\date{\today}

\begin{abstract}
We unveil the relation between the linear Anderson localisation process and nonlinear modulation instability. Anderson localised modes are formed in certain temporal intervals due to the random background noise. Such localised modes seed the formation of solitary waves that will appear during the modulation instability process at those preferred intervals. Afterwards, optical-event horizon effects between dispersive waves and solitons produce an artificial collective acceleration that favours the collision of solitons, which could eventually lead to a rogue-soliton generation. 
\end{abstract}

\maketitle

\section{Introduction} 
Modulation instability (MI) is  one of the most basic and important nonlinear optical processes. In optical fibers, MI induces spectral sidebands to narrowband optical pulses, eventually leading to a broad supercontinuum after a short propagation distance \cite{Coen01,Provino01,Avdokhin03,Schreiber03,Wadsworth04,Vanholsbeeck05,Frosz06}. Generation of supercontinua via long pulses,  lacks coherence and stability in comparison to using ultrashort pulses, mainly because of the amplification of the background noise \cite{Dudley06}. Nevertheless, the understanding of the influence of the input noise on the output spectrum has driven multiple studies \cite{Newbury03,Kobtsev06,Demircan06,Tuurke07}. The evidence presented by Solli \textit{et al.} that the output spectra contain statistically rare rogue events with large intensities and enhanced redshift has  boosted huge research \cite{Solli07,Dudley08,Genty10,Driben12,DeVore13,Dudley14,Armaroli15,SotoCrespo16}, because of the interesting connections between this phenomenon with the rare destructive rogue-waves in oceans \cite{Kharif03,Osborne10}. In optical rogue events, the interplay between nonlinearity and dispersion amplifies the background noise that leads to  pulse break-up. Because of the Raman nonlinearity, solitons are continuously redshifted and decelerate in the time domain \cite{Dianov85,Mitschke86}. Depending on the  initial input noise, a rogue soliton with very large intensity may
appear inside the fibre after multiple soliton-soliton collisions. 

In this article, we present  new groundbreaking details on the dynamics that proceed the collision of solitons inside of optical fibres. In particular, we describe two different processes: solitonisation of Anderson localisation (AL), followed by optical-event-horizon (OEH) induced self-frequency redshift (or deceleration in the time domain). The first process is the key element in generating solitons from the background noise during MI, whereas the collision of different solitons is facilitated by the second process. The reported concepts will  be definitely important for a variety of other related fields, governed by similar models,  such as ocean-wave physics, and Bose-Einstein Condensates.

In 1958 Philip Anderson reported that disorder can induce linear localised states in solid crystals, due to the cumulative effects of multiple scatterings  \cite{Anderson58}. In optics, disorder-induced transverse localisation has been demonstrated in mm-long photonic lattices  \cite{Raedt89,Schwartz07,Lahini08,Lahini09,Szameit10,Levi11,Segev13,Leonetti14a,Leonetti14c}. Ensembling over multiple realisations of disorder was needed, since the propagation distance is too short to provide self-averaging. An important condition for AL to exist is to maintain the disorder along the direction of propagation. The influence of having a fluctuating disorder could result in superdiffussion of the optical beam, known as a hyper-transport effect \cite{Jayannavar82,Golubovic91,Rosenbluth92,Arvedson06,Levi12}. Multiple counter-intuitive  phenomena have also been reported due to the interplay between nonlinearity and AL \cite{Kopidakis00,Schwartz07,Pikovsky08,Conti08,Flach09,Fishman12,Conti12,Dardiry12,Leonetti14b}, yet a robust picture of this complex interaction is still under quest. Here, we show how a nonlinear-induced temporal potential that is disordered and slowly-evolving (quasi-static)   leads to soliton formation during the MI process.

\section{Anderson localisation and modulation instability}
\subsection{Generalised nonlinear Schr\"{o}dinger equation}   
Using the slowly-varying envelope  approximation (SVEA), the propagation of intense pulses in optical fibres can be described in terms of  the generalised nonlinear Schr\"{o}dinger equation \cite{Agrawal07},
\begin{equation}
\left[ i\partial_{z}+\displaystyle\sum_{m\geq 2}\frac{\beta_{m}(i\partial_{t})^{m}}{m!}+\gamma\left( 1+i\tau_{sh}\partial_{t}\right)  \left(R(t)\otimes |A|^{2}\right)\right] A  =0 , 
\label{eq1}
\end{equation}
where  $A\left(z,t\right)$ is the pulse complex envelope,   $t$ is the time-delay in a reference frame moving with the pulse group velocity,  $ \beta_m$, $\gamma $, $\tau_{sh}$ are the dispersion, nonlinear and  self-steepening coefficients, respectively, $R(t)$ is the nonlinear response function that includes  Kerr and Raman contributions,  $\otimes $ denotes the convolution integral, $\gamma =n_{2}\omega_0/ c A_{\mathrm{eff}} $, $n_{2}$ is the nonlinear refractive  in units of m$^2$/W, $\omega_0$ is the pulse central frequency, $c$ is the speed of light in vacuum, and $A_{\mathrm{eff}} $ is the effective optical mode area. Neglecting higher-order nonlinear and dispersion coefficients, Eq. (\ref{eq1}) becomes simply
\begin{equation}
 i\partial_{z}A-\frac{\beta_{2}}{2}\partial_{t}^{2}A+\frac{\omega_0}{c}\Delta n(z,t)A  =0 , 
\label{eq2}
\end{equation}
where the nonlinear-Kerr term results in a pure spatio-temporal modulation of the refractive index $\Delta n(z,t)= n_2 |A|^{2}/A_{\mathrm{eff}} $, which will follow the variation of the pulse intensity during propagation in $z$.  In the anomalous dispersion regime, the amplification of the background noise after a short propagation distance is inevitable for long pulses, resulting in a random temporal modulation of the refractive index. The SVEA  implies that  $\left|\partial_{z} \Delta n \right |\ll\omega_0 \left|\Delta n \right| /c  $,  therefore, $\Delta n(z,t)\approx \Delta n(t)$ is frozen over short spatial $z$-intervals. In other words, Eq. (\ref{eq2}) becomes the 1-D temporal analogue of the transverse-disorder Anderson waveguides, where linear temporal Anderson localised states could be formed. Optical fibres are usually meters long and the complex envelope varies slowly over tenth of centimetres or even meters  (i.e. 1 -- 2 orders of magnitude longer than the waveguides used in AL experiments \cite{Raedt89,Schwartz07,Lahini08,Lahini09,Szameit10,Levi11,Segev13,Leonetti14a,Leonetti14c}). Therefore, temporal Anderson localisation could be realised in optical fibres for any input-noise profile. The localisation is due to the interference between forward and backward waves in the time domain of the moving reference frame that allows positive and negative delays without violating causality. 

\begin{figure*}
\centering
\includegraphics[width=16cm, height=8.6 cm]{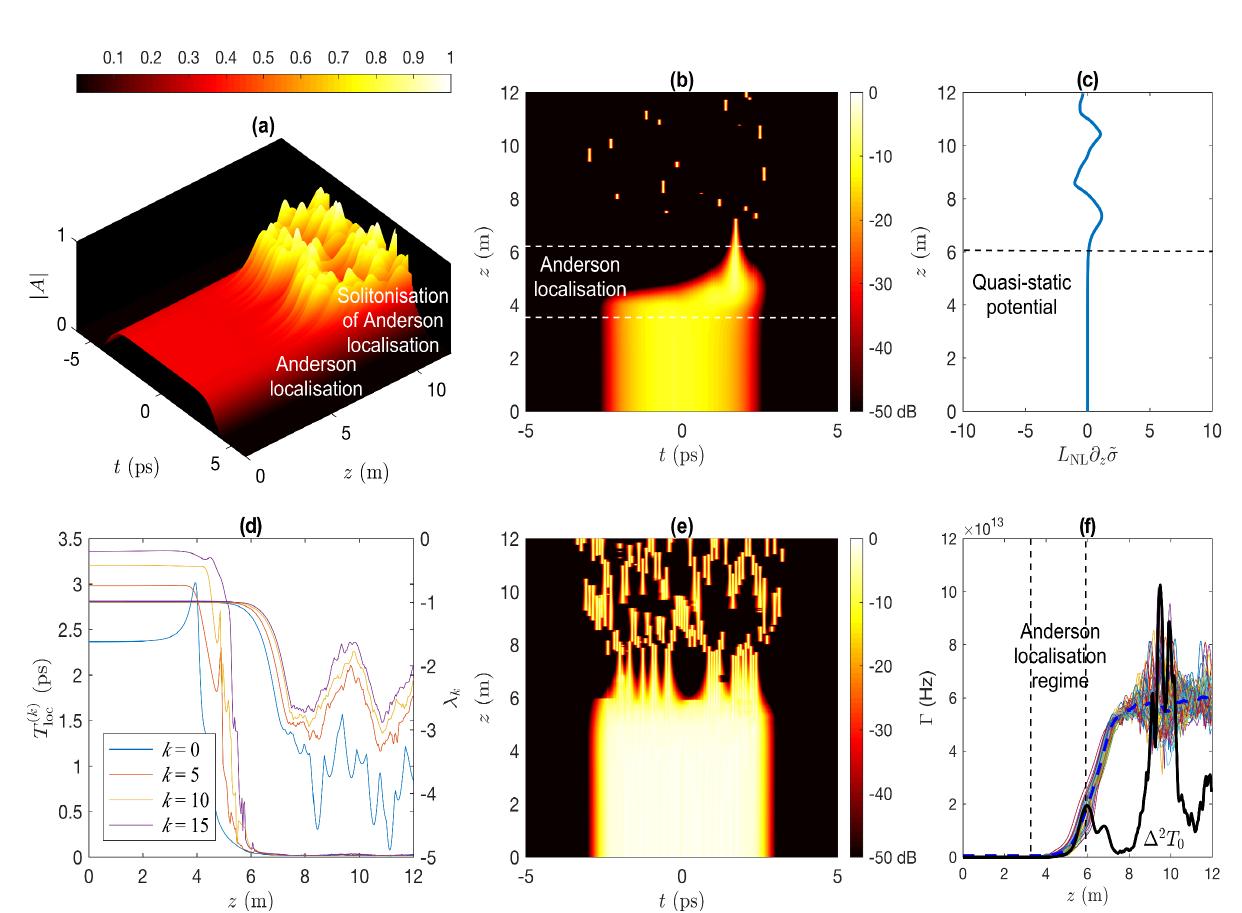}  
\caption{ (a) Temporal evolution of a superGaussian pulse $A=\exp\left[-1/2\:(t/T_0)^{10}\right]$ at wavelength 1060 nm, with $T_0 = $ 3.63 ps (FWHM = 7 ps) and input power 100 W inside the solid silica-core photonic crystal fibre of Ref. \cite{Dudley08} in the absence of higher-order dispersion, Raman effect and self-steepening. (b) Temporal evolution of the ground Anderson state of the induced temporal-waveguide. (c) Spatial dependence of the quantity  $L_{\rm{NL}}\: \partial_{z}\tilde{\sigma}^2 $. (d) Spatial evolution of the  localisation times and eigenvalues of four Anderson modes with $k=0,5,10,15$.  (e) Temporal evolution of the amplitude of the first 20 linear modes on the top of each other. Each mode is normalised such that its energy is unity. (f) Spatial dependency of the Lyapunov exponent $\Gamma$, its mean (dashed blue), and variance (solid black) of an ensemble of 50 different input shot noise.
\label{Fig1}}
\end{figure*}

\subsection{Linear localised Anderson modes in temporal potentials} 
The random temporal modulation of the refractive index induced by strong long pulses during  the propagation inside optical fibres corresponds to an optical potential $U=\omega_0\Delta n/c$ with  allowed linear localised Anderson modes, which will  follow the track of the pulse. These modes are the solutions of the following linear Schr\"{o}dinger equation,
\begin{equation}
 i\partial_{z}u_{k}-\frac{\beta_{2}}{2}\partial_{t}^{2}u_{k}+U(z,t)u_{k}  =0, 
\label{eq4}
\end{equation}
where $u_{k}$ is the complex amplitude of the mode $k=0,1,2,..$. Expressing $u_{k}(z,t)=f_{k}(t)\exp(i\lambda_{k} z)$, Eq. (\ref{eq4}) becomes an eigenvalue problem with $f_{k}$ and $\lambda_{k}$ representing the eigenfunctions and eigenvalues, respectively. The localisation time $T^{(k)}_{\rm loc}$  is given by \cite{Conti12,Lifshitz88},
\begin{equation}
T^{(k)}_{\rm loc}\left(z\right)=\frac{\left(\displaystyle\int \left|f_{k}(z,t)\right|^2 dt\right)^2} {\displaystyle \int\left|f_{k}(z,t)\right|^4 dt}.
\label{eq6}
\end{equation}
For the ground state, $T^{(0)}_{\rm loc}\left(z\right)$ is the inverse  of the Lyapunov exponent $\Gamma$ that is a self-averaging quantity \cite{Limonov12}. $\Gamma$  satisfy the single parameter scaling equation, which  in 1-D random temporal scheme is given by $\Delta^2=\Gamma/T_{0}$, where $\Delta^2 $ is the variance of $\Gamma$ and $T_{0}$ is the input pulse width \cite{Anderson80,Deych00}. We have maintained the spatial dependence of $T^{(k)}_{\rm loc}$ and $\Gamma$ to determine the effect of the SVEA on these quantities.

\subsection{Solitonisation of Anderson localisation} 
Consider the propagation of a long superGaussian pulse with central wavelength $1060$~nm, full-width-half-maximum  (FWHM) $7$~ps, and input power $100$~W in a solid silica-core photonic crystal fibre  with zero-dispersion wavelength located at $1055$~nm. The coefficients of dispersion, Kerr nonlinearity and self-steepening follow Ref.~\cite{Dudley08}. A superGaussian pulse induces a temporal quasi step-index rectangular waveguide. To understand the role of the noise, we first simulate Eqs. (\ref{eq2}-\ref{eq4}). Figure \ref{Fig1}(a) show the temporal evolution of the nonlinear pulse. Starting from $z=  0$, the background noise builds up and then alters the pulse amplitude around $z \approx 6$~m due to MI.  In panel (b), it depicts how the linear fundamental  mode is adiabatically compressed significantly due to AL until  $z\approx 6$ m. The temporal position of localisation is completely random, and it relies on the shape of the input noise. Since the temporal-induced potential is evolving, the localisation process is halted at that position, and the fundamental mode attempts to sustain at other places for few centimetres. Discretising the optical fibre into small segments after MI occurs, one could regard each segment as a different random, where AL attempts to occur.  As a posteriori justification of our assumption of having `quasi-static' random potential, we have plotted in panel (c) the normalised quantity $L_{\rm{NL}}\: \partial_{z}\tilde{\sigma}^2 $, where $\tilde{\sigma}^2=\sigma^2/\sigma^2(0)$,  $\sigma^2$ is the variance of the potential $U$, and $L_{\rm{NL}}$ is the nonlinear length  \cite{Agrawal07}. Clearly, the potential varies vary slightly in comparison to its initial value.

The characteristics of the fundamental Anderson mode measured by its localisation time and eigenvalue are  shown in panel (d). In the regime where AL exists, the temporal width and the eigenvalue of the ground mode are significantly reduced. When AL stops, the fundamental mode start to jump from one potential dip to another, resulting in fluctuating behaviour of its eigenvalue. In panel (e), we show the first twenty  Anderson eigenmodes on the top of each other. A different combination of these modes are randomly excited based on the input pulse. These modes seed the emission of solitons shown in \ref{Fig1}(a),  a  phenomenon known as solitonisation of Anderson localisation \cite{Conti12}.  The characteristics of some of these modes  are plotted in panel (d). Solitons are emitted with close amplitudes around $z\approx 6$  m, because their corresponding modes have close eigenvalues, and  soliton amplitudes are proportional to $\sqrt{\left|\lambda_k\right|}$ \cite{Agrawal07}. We would like to emphasise that this is different from  the traditional case where a regular soliton is always associated with  two  trapped even and odd modes due to the fixed relation between the soliton amplitude and  width \cite{Saleh13}. 

We also studied the dependence of the localisation time of the fundamental mode on the input shot noise. The spatial dependency of $\Gamma$,  its mean, and variance $\Delta^2$ of an ensemble of 50 simulations of different random shot noise are depicted in panel (f). Until $z = 7$~m, $\Gamma$ or $T^{(0)}_{\rm loc}$ is almost independent of the noise input profile. Moreover, $\Gamma$ and $\Delta^2\:T_{0}$ initially follow the single parameter scaling equation until $z\approx 6$  m, where our quasi-static approximation starts to break down, see panel (c).

\section{Optical-event horizons} 
As illustrated in the previous section, solitons are generated after MI with close amplitudes, hence, they are anticipated to follow parallel trajectories that disfavour soliton collisions. Beside AL, optical-event horizons (OEHs) \cite{Philbin08} are also playing a major role in rogue-soliton generation that is mainly based on collision of solitons. The nonlinear temporal waveguide due to a strong pump pulse in a Kerr medium acts as an optical barrier such that a trailing probe pulse with slightly higher group velocity cannot penetrate and bounces back \cite{Philbin08,Wang15}. Because of the cross-phase modulation the probe speed slows down until it matches the strong pulse velocity, so the two pulses are locked together and nonlinearly interact  for a long distance. This phenomenon  has also been predicted for probe powers and  energies closer or even greater than the pump \cite{Demircan13}.  Exploiting this effect, octave-spanning highly-coherent supercontinua  \cite{Demircan13,Demircan14}, and soliton interactions mediated by trapped dispersive waves \cite{Yulin13,Oreshnikov15} can be attained.  Rogue events in the group-velocity horizon due to soliton and dispersive waves interactions have  also been considered \cite{Demircan11,Demircan12}. 

\begin{figure}
\includegraphics[width=8.6cm]{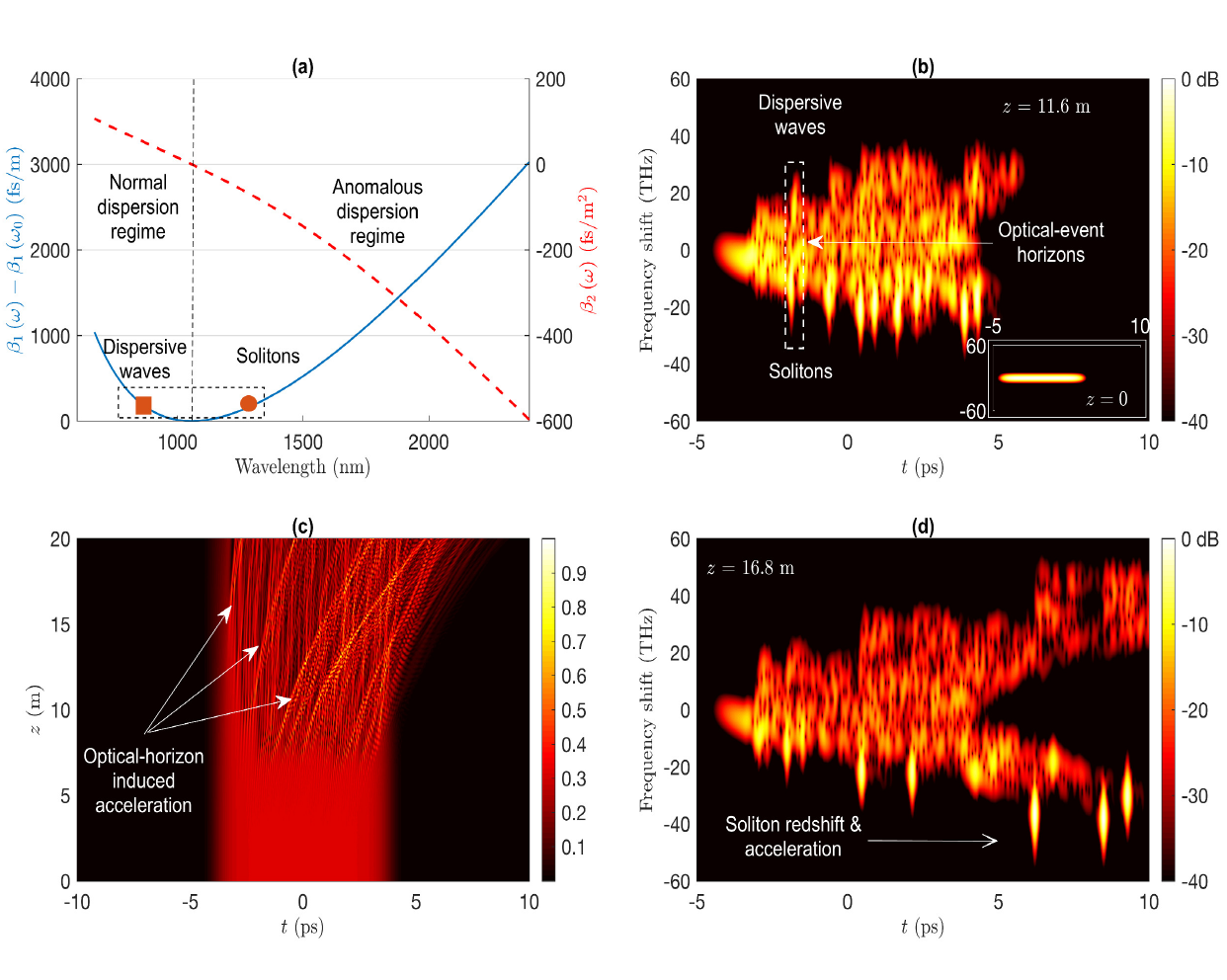}  
\centering
\caption{(a) Wavelength-dependence of first-order $\beta_1$ and second-order $ \beta_2$ dispersion coefficients. The dotted rectangle shows a group of a soliton and a dispersive-wave with nearly group velocities. (b) XFROG representation of the superGaussian pulse  at $z$ = 11.6 m.  The inset is XFROG at $z$ = 0. (c) Temporal evolution of the superGaussian pulse. (d) XFROG representation of the superGaussian pulse  at $z$ = 16.8 m.  Simulations in this figure are performed using Eqs. (\ref{eq1}) and (\ref{eq4}) in the absence of $\beta_{m\geq 4}$, $\tau_{sh}$, and Raman nonlinearity.  \label{Fig2}}
\end{figure}

Adding third-order dispersion in Eq. (\ref{eq2}) will force solitons to emit and trap dispersive waves that are seeded by the background noise, due to satisfying the phase matching conditions.   A large pool of solitons and dispersive waves will form via MI.  Near the zero dispersion wavelength, the concave group-velocity $\beta_1$ dispersion (Fig. \ref{Fig2}(a)) enables synchronous co-propagation of solitons and dispersive waves in the anomalous and normal dispersive regimes, respectively. Hence, the condition for the optical group-velocity event horizon \cite{Philbin08}  between a leading soliton and a trailing dispersive wave can be  easily met. This is evident in the cross frequency optical gating spectrogram (XFROG) at $z$ = 11.6 m in Fig. \ref{Fig2}(b). In this case, the soliton-induced potential barrier impedes the flowing of the dispersive wave and reflect it back after collision. Interestingly, the collision can result in a soliton self-frequency redshift accompanied by a deceleration in the time domain even in the absence of Raman nonlinearity, as depicted in Fig.  \ref{Fig2}(c,d)  that present the temporal evolution of the pulse inside the fibre and the XFROG at $z$ = 16.8 m. This deceleration is stronger for solitons at the leading edge, since they are trailed by a large number dispersive waves. Analytical description of this type of acceleration will be the subject of a future work.  The opposite scenario with trailing soliton colliding with a dispersive wave at slightly different velocity has been exploited for obtaining soliton blueshift \cite{Demircan13,Demircan14}.   

The third-order dispersion enhances the evolution of the background noise starting from $z\approx$ 7  m, after AL takes place. We have found that the inclusion of higher-order dispersion coefficients $\beta_{m>3}$ and self-steepening effects $ \tau_{sh}$ does not involve additional features in pulse dynamics, other than fostering the randomness of the temporal potential. 

\section{Soliton clustering and rogue-soliton generation} 
\begin{figure*}
\centering
\includegraphics[width=16cm, height=9 cm]{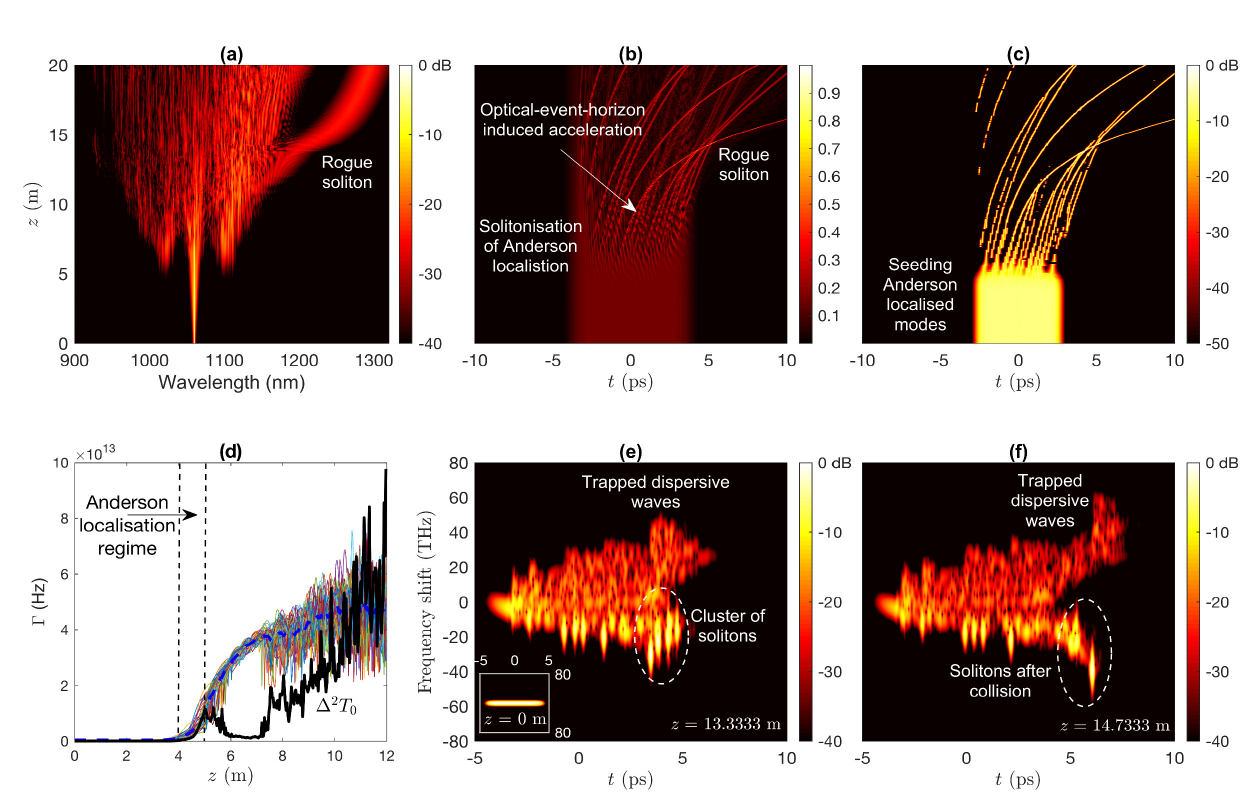}  
\caption{(a,b)  Spectral and temporal evolution of the superGaussian pulse. The temporal contour plot is normalised to its peak. (c) Temporal evolution of the first 20 Anderson eigenmodes. (d) Spatial dependency of the Lyapunov exponent $\Gamma$, its mean (dashed blue), and variance (solid black) of an ensemble of 50 different input shot noise.  (e,f) XFROG representation of the superGaussian pulse inside the fibre at $z=$13.3 m and 14.73 m. The inset in (e) is XFROG at $z=$ 0. Simulations in this figure are performed using the full Eqs. (\ref{eq1}) and (\ref{eq4}). \label{Fig4}}
\end{figure*}

In this section, we will apply our findings of AL and OEHs in elaborating how a rogue soliton is generated. Figure \ref{Fig4}(a,b) displays the spectral and temporal   evolution of the superGaussian pulse with Raman effect producing strong soliton self-frequency redshift and deceleration in the time domain. Simulations in Fig. \ref{Fig4} are performed by solving Eqs. (\ref{eq1}) and (\ref{eq4}), i.e. including Kerr, Raman, self-steepening, and  higher-order dispersion coefficients until the tenth-order. Due to AL, the quasi-static induced random potential allow localised  modes to exist, which in turn seed the emission of  solitary waves as shown in panel (c).  This plot shows the track of the first 20 Anderson modes, which are usually very sufficient to describe the key players in rogue-soliton generation.   The tracks of these modes will experience discontinuities with any slight perturbation, because they have close eigenvalues.  The dependence of the Lyapunov exponent $\Gamma$, its mean, and variance on the input shot noise is presented in panel (d), anticipating the regime where AL takes place. In comparison to the ideal case presented in  \ref{Fig1}(f), the length of tis regime has been reduced to around 1 m long due to the inclusion of higher-order dispersion coefficients, self-steepening, and Raman nonlinearity.

Soliton collisions occur when their trajectories intersect, however, specific scenarios lead to rogue solitons with large amplitude and frequency redshift.  Raman nonlinearity induces soliton acceleration that is proportional to the quartic  of its amplitude. Hence, soliton collisions are  unlikely to initially occur, since solitons will follow nearly  parallel trajectories. However, because of OEH-induced  acceleration due to   solitary and dispersive waves collisions, soliton trajectories start to intersect. Raman nonlinearity allows the solitons to cluster in the time and frequency domains very quickly, as depicted in the XFROG representation in Fig. \ref{Fig4}(e,f). This results in strong temporal overlap and close group velocities for the solitons, so they could strongly nonlinearly interact for a long distance and a rogue-soliton is generated after  exchanging energy between them, as shown in panels (b,f). This cluster of solitons is analogous to an OEH made of different pulses, however, with similar amplitudes.

The statistics of a rogue-soliton emission that follows the L-shaped Weibull distribution depends mainly on the intense of soliton-soliton collisions and the fibre length. In the case of pumping near the zero-dispersion wavelength, soliton collisions are nearly inevitable due to OEH-induced acceleration that changes the soliton paths. We have found only around 50 events, where solitons could have parallel trajectories inside a fibre of 20 m long, from an ensemble of 1000 different input noise. This number is also significantly dropped if the fibre length is increased.  The intensity of a rogue-soliton is based on how many clusters are formed,  how many solitons are within each cluster, and how these solitons are temporally overlapped.

\section{Conclusions} 
We have reported the missing ingredients of the generation of a rogue-soliton in optical fibres. We have found that the true origin of soliton generation during the modulation instability process is the temporal  Anderson localisation effect. Solitary-waves that appear from nowhere during propagation of a nonlinear pulse inside an optical fibre are seeded by the linear eigenmodes of a 1-D temporal Anderson crystal. The usual pumping near the zero-dispersion-wavelength for obtaining  broadband supercontinua favours optical-event horizons formation between solitons and dispersive waves. The latter effect results in an additional soliton acceleration, besides the usual Raman-induced one. This acceleration is unbalanced towards the pulse leading edge and favours soliton-soliton collisions. Rogue solitons are generated after strong temporal overlap between individual solitons with close group velocities, due to Raman nonlinearity.

We believe that our findings solve one of the most debated questions in the field, namely: are rogue waves generated by processes that are intrinsically linear or nonlinear? The answer to this question is that a linear  process, Anderson localisation, seeds temporal localised structures from the background noise. Subsequently a nonlinear phenomenon, the solitonisation of Anderson localisation, together with optical-event horizons induce the last steps of rogue-soliton formation.

These rich mechanisms demonstrate the complexity underlying rogue-soliton generation and furnish a clear evidence of the previously-unconsidered temporal Anderson localisation, and the unexpected interaction with optical horizons. This scenario will potentially lead to novel routes for controlling of extreme nonlinear waves via linear-disorder optimisation.

\section*{Funding Information}
Royal Society of Edinburgh.


\end{document}